\documentclass[12pt,letterpaper]{article}
\usepackage[algoruled,linesnumbered,algo2e,vlined]{algorithm2e}
\usepackage[draft,mode=multiuser]{fixme}
\usepackage{amsmath}
\usepackage{times}
\usepackage{pifont}
\usepackage{amsmath,amssymb}
\usepackage{url}
\newcommand{\cmark}{\ding{51}}
\usepackage{rotating}

\fxuseenvlayout{colorsig}
\fxusetargetlayout{color}

\FXRegisterAuthor{kg}{akg}{KG}
\FXRegisterAuthor{hb}{ahb}{HB}
\definecolor{kgnote}{rgb}{1.0000,0.0000,0.0000}
\fxsetface{env}{}

\usepackage{Latex-ieee6x9}

\newcommand{\Prefix}{\mathit{prefix}}
\newcommand{\lcp}{\mathit{lcp}}

\newcommand{\LPF}{\mathit{LPF}}
\newcommand{\PrevOcc}{\mathit{PrevOcc}}

\newcommand{\PSVN}{\mathit{PSV}}
\newcommand{\NSVN}{\mathit{NSV}}
\newcommand{\PSV}{\mathit{PSV_{\mathit{lex}}}}
\newcommand{\NSV}{\mathit{NSV_{\mathit{lex}}}}
\newcommand{\PSVT}{\mathit{PSV_{\mathit{text}}}}
\newcommand{\NSVT}{\mathit{NSV_{\mathit{text}}}}

\newcommand{\PNSV}{\mathit{PNSV}}
\newcommand{\SA}{\mathit{SA}}
\newcommand{\RANK}{\mathit{SA}^{-1}}
\newcommand{\ignore}[1]{}
\newtheorem{definition}{Definition}

\title{Simpler and Faster  Lempel Ziv Factorization}
\author{Keisuke Goto and Hideo Bannai\\
{\small Department of Informatics, Kyushu University, Fukuoka
  819-0395, Japan}\\
{\small {\tt \{keisuke.gotou,bannai\}@inf.kyushu-u.ac.jp}}
}

\begin{document}
\maketitle
\begin{abstract}
  We present a new, simple, and efficient approach
  for computing the Lempel-Ziv (LZ77) factorization of a string in
  linear time, based on suffix arrays.
  Computational experiments on various data sets show that our
  approach constantly outperforms the fastest previous algorithm
  LZ\_OG (Ohlebusch and Gog 2011),
  and can be up to 2 to 3 times faster
  in the processing after obtaining the suffix array,
  while requiring the same or a little more space.
\end{abstract}

\section{Introduction}
The LZ77 factorization~\cite{LZ77} of a string
captures important properties concerning repeated occurrences
of substrings in the string,
and has obvious applications in the field of data compression,
as well as being the key component
to various efficient algorithms on strings~\cite{kolpakov99:_findin_maxim_repet_in_word,duval04:_linear}.
Consequently, many algorithms for its efficient calculation have been
proposed.
The LZ77 factorization of a string $S$ is a factorization
$S = f_1 \cdots f_n$ where each factor $f_k$ is either 
(1) a single character if that character does not occur in $f_1\cdots
f_{k-1}$, or,
(2) the longest prefix of the rest of the string which occurs 
at least twice in $f_1\cdots f_{k}$.

A na\"ive algorithm that computes the longest common prefix
with each of the $O(N)$ previous positions only requires
$O(1)$ working space (excluding the output),
but can take $O(N^2)$ time, where $N$ is the length of the string.
Using string indicies such as suffix trees~\cite{Weiner}
and on-line algorithms to construct them~\cite{Ukk95},
the LZ factorization can be computed in an on-line
manner in $O(N\log|\Sigma|)$ time and $O(N)$ space,
where $|\Sigma|$ is the size of the alphabet.

Most recent efficient linear time algorithms are off-line,
running in $O(N)$ time for integer alphabets using $O(N)$ 
space (See Table~\ref{table:lineartimealgorithms}).
They first construct the suffix array~\cite{manber93:_suffix} of the string,
and compute an array called the Longest Previous Factor (LPF) array
from which the LZ factorization can be easily
computed~\cite{crochemore08:_simpl_algor_comput_lempel_ziv_factor,chen08:_lempel_ziv_factor_using_less_time_space,crochemore09:_lpf_comput_revis,ohlebusch11:_lempel_ziv_factor_revis,a.ss:_lempel_ziv_lz77}.
Many algorithms of this family first compute the 
longest common prefix (LCP) array prior to the computation of the LPF array.
However, the computation of the LCP array is also costly.
The algorithm CI1 (COMPUTE\_LPF)
of~\cite{crochemore08:_comput_longes_previous_factor},
and the algorithm
LZ\_OG~\cite{ohlebusch11:_lempel_ziv_factor_revis}
cleverly avoids its computation and directly computes the LPF array.

An important observation here is that the LPF
is actually more
information than is required for the
computation of the LZ factorization, i.e.,
if our objective is the LZ factorization, we only use
a subset of the entries in the LPF
.
However, the above algorithms focus on computing the entire LPF array,
perhaps since it is difficult to determine beforehand, 
which entries of LPF are actually required.
Although some algorithms such as
a variant of CPS1~\cite{chen08:_lempel_ziv_factor_using_less_time_space}
or CPS2 in~\cite{chen08:_lempel_ziv_factor_using_less_time_space}
avoid computation of LPF,
they either require the LCP array, or do not run in linear worst case time
and are not as efficient. (See ~\cite{a.ss:_lempel_ziv_lz77} for a survey.)

In this paper, we propose a new approach to avoid the computation of
LCP and LPF arrays altogether,
by combining the ideas of the na\"ive algorithm with those of CI1 and LZ\_OG,
and still achieve worst case linear time.
The resulting algorithm is surprisingly both simple and efficient.

Computational experiments on various data sets shows that
our algorithm constantly outperforms
 LZ\_OG~\cite{ohlebusch11:_lempel_ziv_factor_revis},
and can be up to 2 to 3 times faster
in the processing after obtaining the suffix array,
while requiring the same or a little more space.

Although our algorithm might be considered as a simple combination of ideas
appearing in previous works, this paper is one of the first to 
propose, implement and evaluate this combination.
We note that algorithms that avoid the computation of LCP and LPF
based on similar ideas as in this paper
were developed independently and almost simultaneously
by Kempa and Puglisi~\cite{kempa_lz} and K\"{a}rkk\"{a}inen et
al.~\cite{karkkainen_lz}.
Since we did not have knowledge of their work until very recently,
we have not made comparisons between them.
The worst case time complexity of~\cite{kempa_lz} is
not independent of alphabet size, but is fast and space efficient. 
In the more recent manuscript~\cite{karkkainen_lz},
two new linear time algorithms which outperform 
all previous algorithms (including ours) in terms of time and space
are proposed, asserting the potential of this approach.

\begin{table}[t]
  \begin{center}
    \caption{Fast Linear time LZ-Factorization Algorithms based on
      Suffix Arrays}
    \label{table:lineartimealgorithms}
    \begin{tabular}{|l|c|c|c|c|}\hline
      algorithm & worst case time & SA & LCP
      & LPF, PrevOcc\\\hline
      CI1~\cite{crochemore08:_comput_longes_previous_factor} &
      $\Theta(N)$ & \cmark & & \cmark\\
      CI2~\cite{crochemore08:_comput_longes_previous_factor} &
      $\Theta(N)$ & \cmark & \cmark & \cmark\\
      CPS1~\cite{chen08:_lempel_ziv_factor_using_less_time_space} &
      $\Theta(N)$ & \cmark & \cmark & \\ %%& &\\
      CIS~\cite{crochemore08:_simpl_algor_comput_lempel_ziv_factor} &
      $\Theta(N)$ & \cmark & \cmark & \cmark \\%%& &\\
      CII~\cite{crochemore09:_lpf_comput_revis} &
      $\Theta(N)$ & \cmark & \cmark & \cmark \\%%& &\\
      LZ\_OG~\cite{ohlebusch11:_lempel_ziv_factor_revis} & $\Theta(N)$
      & \cmark & & \cmark\\
      this work, \cite{karkkainen_lz} & $\Theta(N)$ & \cmark &     &  \\\hline
      Naive & $\Theta(N^2)$ &&&\\\hline %%&& $O(1)$\\\hline

    \end{tabular}
  \end{center}
\end{table}

\section{Preliminaries}

Let $\mathcal{N}$ be the set of non-negative integers.
Let $\Sigma$ be a finite {\em alphabet}.
An element of $\Sigma^*$ is called a {\em string}.
The length of a string $T$ is denoted by $|T|$. 
The empty string $\varepsilon$ is the string of length 0,
namely, $|\varepsilon| = 0$.
Let $\Sigma^+ = \Sigma^* - \{\varepsilon\}$.
For a string $S = XYZ$, $X$, $Y$ and $Z$ are called
a \emph{prefix}, \emph{substring}, and \emph{suffix} of $T$, respectively.
The set of prefixes of $T$ is denoted by $\Prefix(T)$.
The \emph{longest common prefix} of strings $X,Y$, denoted $\lcp(X, Y)$, 
is the longest string in $\Prefix(X) \cap \Prefix(Y)$.

The $i$-th character of a string $T$ is denoted by 
$T[i]$ for $1 \leq i \leq |T|$,
and the substring of a string $T$ that begins at position $i$ and
ends at position $j$ is denoted by $T[i..j]$ for $1 \leq i \leq j \leq |T|$.
For convenience, let $T[i..j] = \varepsilon$ if $j < i$,
and $T[|T|+1] = \$$ where $\$$ is a special delimiter character that
does not occur elsewhere in the string.

\subsection{Suffix Arrays}
The suffix array~\cite{manber93:_suffix} $\SA$ of any string $T$
is an array of length $|T|$ such that
for any $1 \leq i \leq |T|$,
$\SA[i] = j$ indicates that $T[j:|T|]$ is the $i$-th lexicographically smallest suffix of $T$.
For convenience, assume that $\SA[0] = |T|+1$.
The inverse array $\RANK$ of $\SA$ is an array of length $|T|$ such that
$\RANK[\SA[i]] = i$.
As in~\cite{karkkainen09_plcp}, let $\Phi$ be an array of length $|T|$ such that
$\Phi[\SA[1]] = |T|$ and
$\Phi[\SA[i]] = \SA[i-1]$ for $2 \leq i \leq |T|$,
i.e., for any suffix $j = \SA[i]$,
$\Phi[j] = \SA[i-1]$ is the immediately preceding suffix in
the suffix array.
The suffix array $\SA$ for any string of length $|T|$
can be constructed in $O(|T|)$ 
time
regardless of the alphabet size, assuming an integer
alphabet~(e.g.~\cite{Karkkainen_Sanders_icalp03}).
All our algorithms will assume that the $\SA$ is already computed.
Given $\SA$, arrays $\RANK$ and $\Phi$ can easily be computed in
linear time by a simple scan.

\subsection{LZ Encodings}
LZ encodings are dynamic dictionary based encodings with many variants.
The variant we consider is also known as the s-factorization~\cite{crochemore84:_linear}.

\begin{definition}[LZ77-factorization]
  \label{def:s_factorization}
  The s-factorization of a string $T$ is
  the factorization $T = f_1 \cdots f_n$ where each
  s-factor $f_k\in\Sigma^+~(k=1,\ldots,n)$
  is defined inductively as follows:
  $f_1 = T[1]$. For $k \geq 2$:
  if $T[|f_1\cdots f_{k-1}|+1] = c \in \Sigma$ does not occur in $f_1\cdots f_{k-1}$,
  then $f_k = c$. Otherwise, $f_k$ is the longest prefix of $f_k
  \cdots f_n$ that occurs at least twice in $f_1 \cdots f_k$.
\end{definition}
Note that each LZ factor can be represented in constant space,
i.e., 
a pair of integers where the first and second elements
respectively represent the length and position of 
a previous occurrence of the factor.
If the factor is a new character and the length of its previous
occurrence is $0$, the second element will encode the new character
instead of the position.
For example the s-factorization of the string
  $T = \mathtt{abaabababaaaaabbabab}$ is
  $\mathtt{a}$,
  $\mathtt{b}$,
  $\mathtt{a}$,
  $\mathtt{aba}$,
  $\mathtt{baba}$,
  $\mathtt{aaaa}$,
  $\mathtt{b}$,
  $\mathtt{babab}$. This can be represented as
  $(0, \mathtt{a})$, $(0, \mathtt{b})$, $(1,1)$, $(3,1)$,
  $(4,5)$, $(4,10)$, $(1,2)$, $(5,5)$.

We define two functions $\LPF$ and $\PrevOcc$ below.
For any $1\leq i \leq N$,
$\LPF(i)$ is the longest length of longest common prefix
between $T[i:N]$ and $T[j:N]$ for any $1 \leq j < i$, and
$\PrevOcc(i)$ is a position $j$ which achieves gives
$\LPF(i)$\footnote{There can be multiple choices of $j$, but
  here, it suffices to fix one.}.
More precisely,
\begin{eqnarray*}
\LPF(i) &=& \max(\{0\}\cup\{\lcp(T[i:N],T[j:N]) \mid 1 \leq j < i \})\\
\mbox{and}\\
\PrevOcc(i) &=& 
\begin{cases}
  -1      & \mbox{if } \LPF(i) = 0\\
  j       & \mbox{otherwise}
\end{cases}
\end{eqnarray*}
where $j$ satisfies $1 \leq j < i$, and $T[i:i+\LPF(i)-1] = T[j:j+\LPF(i)-1]$.
Let $p_k = |f_1 \cdots f_{k-1}|+1$.
Then, $f_k$ can be represented as a pair
$(\LPF(p_k), \PrevOcc(p_k))$ if $\LPF(p_k)>0$,
and $(0, T[p_k])$ otherwise.

Most recent fast linear time algorithms for computing the LZ
factorization calculate $\LPF$ and $\PrevOcc$ for all positions
$1\leq i \leq N$ of the text and store the values in an array,
and then use these values as in
Algorithm~\ref{algo:lz_from_lpf} to output the LZ factorization.
\begin{algorithm2e}[t]
  \caption{LZ Factorization from $\LPF$ and $\PrevOcc$ arrays}
  \label{algo:lz_from_lpf}
  \SetKwInOut{Input}{Input}\SetKwInOut{Output}{Output}
  \Input{String $T$, $\LPF$, $\PrevOcc$}
  $p \leftarrow 1$\;
  \While{$p \leq N$}{
    \lIf{$\LPF[p] = 0$}{%
      \Output{$(1, T[p])$}
    }\lElse{%
      \Output{$(\LPF[p], \PrevOcc[p])$}
    }
    $p \leftarrow p + \max(1, \LPF[p])$\;
  }
  
\end{algorithm2e}

\section{Algorithm}
\label{sec:algorithm}
We first describe the na\"ive algorithm for calculating the LZ
factorization of a string, and analyze its time complexity
The na\"ive algorithm does not compute all values of
$\LPF$ and $\PrevOcc$ as explicit arrays,
but only the values required to represent each factor.
The procedure is shown in Algorithm~\ref{algo:naive}.
For a factor starting at position $p$, the algorithm computes
$\LPF(p)$ and $\PrevOcc(p)$ by simply looking at each of its $p-1$
previous positions,
and naively computes the longest common prefix (lcp) between
each previous suffix and the suffix starting at position $p$,
and outputs the factor accordingly.
At first glance, this algorithm looks like an $O(N^3)$ time
algorithm since there are 3 nested loops.
However, the total time can be bounded by $O(N^2)$,
since the total length of the longest lcp's found for each $p$ in the algorithm,
i.e., the total length of the LZ factors found, is $N$.
More precisely, 
let the LZ factorization of string $T$ of length $N$ be $f_1\cdots f_n$, and
$p_k = |f_1 \cdots f_{k-1}|+1$ as before. Then,
the number of character comparisons executed in Line~\ref{algo:naive_charcomp} of Algorithm~\ref{algo:naive}
when calculating $f_k$ is at most $(p_k-1) |f_k+1|$, and the total can be
bounded:
$\sum_{k=1}^n (p_k-1)|f_k+1| \leq N\sum_{k=1}^n|f_k+1| = O(N^2)$.
An important observation here is that if we can somehow reduce the
number of previous candidate
positions for na\"ively computing lcp's (i.e. the choice of $j$ in
Line~\ref{algo:naive_choices} of Algorithm~\ref{algo:naive})
from $O(N)$ to $O(1)$ positions,
this would result in a $O(N)$ time algorithm.
This very simple observation is the first key to the linear running times of
our new algorithms.

\begin{algorithm2e}[t]
  \caption{Na\"ive Algorithm for Calculating LZ factorization}
  \label{algo:naive}
  \SetKwInOut{Input}{Input}\SetKwInOut{Output}{Output}
  \Input{String $T$}
  $p \leftarrow 1$\;
  \While{$p \leq |T|$}{
    $\LPF \leftarrow 0$\;
    \For{$j \leftarrow 1, \ldots, p-1$}{\label{algo:naive_choices}
      $l \leftarrow 0$\;
      \lWhile{$T[j+l] = T[p+l]$\label{algo:naive_charcomp}}{ $l
        \leftarrow l + 1$\tcp*{$l \leftarrow \lcp(T[j:N],T[p:N])$} }
      \lIf{$l > \LPF$}{ $\LPF \leftarrow l$; $\PrevOcc \leftarrow j$\; }
    }
    \lIf{$\LPF > 0$}{\Output{$(\LPF, \PrevOcc)$}}
    \lElse{\Output{$(0, T[p])$}}
    $p \leftarrow p + \max(1, \LPF)$\;
  }
\end{algorithm2e}

To accomplish this, our algorithm utilizes yet another simple but
key observation made in~\cite{crochemore08:_comput_longes_previous_factor}.
Since suffixes in the suffix arrays are lexicographically sorted,
if we fix a suffix $\SA[i]$ in the suffix array,
we know that suffixes appearing closer in the suffix array will have
longer longest common prefixes with suffix $\SA[i]$.

For any position $1 \leq i \leq N$ of the suffix array,
let
\begin{eqnarray*}
  \PSV[i] &=& \max(\{0\}\cup \{ 1 \leq j < i \mid \SA[j] < \SA[i]\})\\
  \NSV[i] &=& \min(\{0\}\cup \{ N \geq j > i \mid \SA[j] < \SA[i] \})
\end{eqnarray*}
i.e., for the suffix starting at text position $\SA[i]$,
the values $\PSV[i]$ and $\NSV[i]$ represent the lexicographic rank
of the suffixes that start before it in the string
and are lexicographically closest (previous and next) to it,
or $0$ if such a suffix does not exist.
From the above arguments, we have that for any
text position $1 \leq p \leq N$,
\begin{eqnarray*}
  \LPF(p) = 
  \max(
  \lcp(T[\SA[\PSV[\RANK[p]]]:N],T[p:N]),\\
  \lcp(T[\SA[\NSV[\RANK[p]]]:N],T[p:N])
  ).
\end{eqnarray*}

The above observation or its variant has been used
as the basis for calculating $\LPF(i)$ for all
$1 \leq i \leq N$ in linear time in practically all previous linear
time algorithms for LZ factorization based on the
suffix array.
In~\cite{ohlebusch11:_lempel_ziv_factor_revis}, they consider
(implicitly) the arrays in text order rather than lexicographic order.
In this case,
\begin{eqnarray*}
  \PSVT[\SA[i]] &=& \SA[\PSV[i]]\\
  \NSVT[\SA[i]] &=& \SA[\NSV[i]]
\end{eqnarray*}
and therefore
\begin{eqnarray*}
  \LPF(p) = 
  \max(
  \lcp(T[\PSVT[p]]:N],T[p:N]),
  \lcp(T[\NSVT[p]]:N],T[p:N])
  ).
\end{eqnarray*}

While~\cite{crochemore08:_comput_longes_previous_factor}
and~\cite{ohlebusch11:_lempel_ziv_factor_revis} utilize this
observation to compute all entries of $\LPF$ in linear time,
we utilize it in a slightly different way
as mentioned previously, and use it to reduce the candidate positions for
calculating $\PrevOcc(i)$ (i.e. the choice of $j$ in Algorithm~\ref{algo:naive})
to only 2 positions. 
The key idea of our approach is in the combination of the above observation
with the amortized analysis of the na\"ive algorithm, suggesting
that we can defer the computation of the values of $\LPF$ until we
actually require them for the LZ factorization and still achieve
linear worst case time.
If $\PSV[i]$ and $\NSV[i]$ (or $\PSVT[i]$ and
$\NSVT[i]$) are known for
all $1 \leq i \leq N$, the linear running
time of the algorithm follows from the previous arguments.
The basic structure of our algorithm is shown in
Algorithm~\ref{algo:basic} when using $\PSV$ and $\NSV$.
Note that it is easy to replace them with $\PSVT$ and $\NSVT$,
and in such case, $\SA$ and $\RANK$ are not necessary once we have
$\PSVT$ and $\NSVT$.

What remains is how to compute
$\PSV[i]$ and $\NSV[i]$,
or,
$\PSVT[i]$ and $\NSVT[i]$
for all $1\leq
i\leq N$.
This can be done in several ways. We consider 3 variations.

The first is a computation of $\PSV[i]$, $\NSV[i]$ using a simple
linear time scan of the suffix array with the help of a stack.
The procedure is shown in Algorithm~\ref{algo:psvnsv}.
This variant requires the text, and the arrays $\SA$, $\RANK$,
$\PSV$, $\NSV$ and a stack. The total space complexity is
$17N + 4S_{max}$ bytes assuming that an integer occupies 4 bytes,
where $S_{max}$ is the maximum size of the stack during the execution
of the algorithm and can be $\Theta(n)$ in the worstcase. We will call this variant BGS.

The other two is a process called {\em peak elimination},
which is very briefly described
in~\cite{crochemore08:_comput_longes_previous_factor}
for lexicographic order (Shown in Algorithms~\ref{algo:psvnsv_ci} and~\ref{algo:peakElimLex}),
and in~\cite{ohlebusch11:_lempel_ziv_factor_revis}
for text order (Shown in Algorithms~\ref{algo:psvnsv_phi} and~\ref{algo:peakElimText}).
In peak elimination, each suffix $i$ and its lexicographically preceding
suffix $j$ ($\RANK[j] + 1 = \RANK[i]$) is examined in some order
of $i$ (lexicographic or text order).
For simplicity, we only briefly explain the approach for text order.
If $i > j$, this means that
$\PSVT[i] = j$ and if $i < j$, $\NSVT[j] = i$.
When both values of $\PSVT[i]$ and $\NSVT[i]$ are determined, 
$i$ is identified as a peak.
Given a peak $i$, it is possible to {\em eliminate} it,
and determine the value of
either $\NSVT[\PSVT[i]]$ (which will be $\NSVT[i]$ if $\PSVT[i] > \NSVT[i]$) 
or $\PSVT[\NSVT[i]]$ (which will be $\PSVT[i]$ if if $\PSVT[i] < \NSVT[i]$)),
and this process is repeated. The algorithm runs
in linear time since each position can be eliminated only once.
The procedure for lexicographic order is a bit simpler since the 
lexicographic order of calculation implies that $\PSV[i]$ will always
be determined before $\NSV[i]$.

The algorithm of~\cite{ohlebusch11:_lempel_ziv_factor_revis} actually
computes the arrays $\LPF$ and $\PrevOcc$ directly without computing
$\PSVT$ and $\NSVT$.
The algorithm we show is actually a simplification,
deferring the computation of $\LPF$ and $\PrevOcc$,
computing $\PSVT$ and $\NSVT$ instead.

For lexicographic order, we need the text and the arrays
$\SA$, $\RANK$, $\PSV$, $\NSV$ and no stack, giving an algorithm with
$17N$ bytes of working space. We will call this variant BGL.
For text order, although the $\Phi$
array is introduced instead of the $\RANK$ array,
the suffix array is not required after its computation.
Therefore, by reusing the space of $\SA$ for $\PSVT$, the total space
complexity can be reduced to $13N$ bytes. We will call this variant BGT.
Note that although $peakElim_{lex}$ and $peakElim_{text}$
are shown as recursive functions for simplicity,
they are tail recursive and thus can be optimized as loops and will not
require extra space on the call stack.

\subsection{Interleaving $\PSVN$ and $\NSVN$}
\label{subsec:opt_interleave}
Since accesses to $\PSVN$ and $\NSVN$ occur at the same 
or close indices,
it is possible to improve the memory locality of accesses by interleaving the 
values of $\PSVN$ and $\NSVN$, maintaining them in a single array as follows.
Let
$\PNSV$ be an array of length $2N$, and for each position
$1 \leq i \leq 2N$,
$\PNSV[i] = \PSVN[j]$ if $i \mod 2 \equiv 0$, $\NSVN[j]$ otherwise,
where $j = \lfloor i/2 \rfloor$.
Naturally, for any $1 \leq i\leq N$, $\PSVN$ and $\NSVN$ can be accessed as
$\PSVN[i] = \PNSV[2i]$ and
$\NSVN[i] = \PNSV[2i+1]$.
This interleaving can be done for both lexicographic order and
text order.
We will call the variants of our algorithms that incorporate this
optimization, iBGS, iBGL, iBGT.

\begin{algorithm2e}[t]
  \caption{Basic Structure of our Algorithms.}
  \label{algo:basic}
  \SetKwInOut{Input}{Input}\SetKwInOut{Output}{Output}
  \Input{String $T$}
  Calculate $\PSV[i]$ and $\NSV[i]$ for all $i = 1 ... N$\;
  $p \leftarrow 1$\;
  \While{$p \leq N$}{
    $\LPF \leftarrow 0$\;
    \For{$j\in \{\SA[\PSV[\RANK[p]]], \SA[\NSV[\RANK[p]]]\}$}{
      $l \leftarrow 0$\;
      \lWhile{$T[j+l] = T[p+l]$}{ 
        $l \leftarrow l + 1$\tcp*{$l \leftarrow \lcp(T[j:N],T[p:N])$} }
      \lIf{$l > \LPF$}{ $\LPF \leftarrow l$; $\PrevOcc \leftarrow
        j$\; }
    }
    \lIf{$\LPF > 0$}{\Output{$(\LPF, \PrevOcc)$}}
    \lElse{\Output{$(0, T[p])$}}
    $p \leftarrow p + \max(1, \LPF)$\;
  }
\end{algorithm2e}

\begin{algorithm2e}[t]
  \caption{Calculating $\PSV$ and $\NSV$ from $\SA$}
  \label{algo:psvnsv}
  \SetKwInOut{Input}{Input}\SetKwInOut{Output}{Output}
  \Input{Suffix array $\SA$}
  \Output{$\PSV$, $\NSV$}
  Let $S$ be an empty stack\;
  \For{$i \leftarrow 1$ to $N$}{
    $x \leftarrow \SA[i]$\;
     \While{$(\mathbf{not}~S.\mathit{empty}())~\mathbf{and}~(\SA[S.\mathit{top}()] > x)$\label{line:stack_sansyou}}{
      $\NSV[S.\mathit{top}()] \leftarrow i$; $S.\mathit{pop}()$ \label{line:nsv_insert1}\;
    }
    $\PSV[i] \leftarrow$ \lIf{$S.empty()$}{$0$ }\lElse{$S.top()$} \label{line:psv_insert}\;
    $S.push(i)$\label{line:stack_insert}\;
  }
  \While{$\mathbf{not}~S.\mathit{empty}()$}{
    $\NSV[S.top()] \leftarrow 0$; $S.pop()$ \label{line:nsv_insert2}\;
  }
\end{algorithm2e}

\begin{algorithm2e}[t]
  \caption{Calculating $\PSV$ and $\NSV$ from $\SA$ by Peak Elimination.}
  \label{algo:psvnsv_ci}
  \SetKwInOut{Input}{Input}\SetKwInOut{Output}{Output}
  \Input{Suffix array $\SA$}
  \lFor{$i \leftarrow 1$ to $N$}{
    $\NSV[i] \leftarrow 0$\;
  }
  $\PSV[1] \leftarrow 0$\;
  \lFor{$i \leftarrow 2$ to $N$}{
    $peakElim_{lex}(i-1, i)$\;
  }
\end{algorithm2e}

\begin{algorithm2e}[t]
  \caption{Peak Elimination $peakElim_{lex}(j, i)$ in Lexicographic Order.}
  \label{algo:peakElimLex}
  \If{$j = 0$ {\bf or} $\SA[j] < \SA[i]$}{
    $\PSV[i] \leftarrow j$\;
  }\Else(\tcp*[h]{$j \geq 1$ {\bf and} $\SA[j] > \SA[i]$}){
    $\NSV[j] \leftarrow i$\;
    $peakElim_{lex}(\PSV[j],i)$ \tcp*{$j$ was peak.}
  }
\end{algorithm2e}

\begin{algorithm2e}[t]
  \caption{Calculating $\PSVT$ and $\NSVT$ from $\SA$ using $\Phi$.}
  \label{algo:psvnsv_phi}
  \SetKwInOut{Input}{Input}\SetKwInOut{Output}{Output}
  \Input{Suffix array $\SA$}
  $\Phi[\SA[1]] \leftarrow N$\;
  \lFor{$i \leftarrow 2$ to $N$}{ $\Phi[\SA[i]] \leftarrow \SA[i-1]$\; }
  \For{$i \leftarrow 1$ to $N$}{
    $\PSVT[i] \leftarrow \bot$; $\NSVT[i] \leftarrow \bot$\;
  }
  \lFor{$i \leftarrow 1$ to $N$}{
    $peakElim_{text}(\Phi[i], i)$\;
  }
\end{algorithm2e}

\begin{algorithm2e}[t]
  \caption{Peak Elimination $peakElim_{text}(j, i)$}
  \label{algo:peakElimText}
  \If{$j < i$}{
    $\PSVT[i] \leftarrow j$\;
    \lIf{$\NSVT[i] \neq \bot$}{
      $peakElim_{text}(j,\NSVT[i])$ \tcp*{$i$ was peak.}
    }
  }
  \Else(\tcp*[h]{$j > i$}){
    $\NSVT[j] \leftarrow i$\;
    \lIf{$\PSVT[j] \neq \bot$}{
      $peakElim_{text}(\PSVT[j], i)$ \tcp*{$j$ was peak.}
    }
  }
\end{algorithm2e}

\section{Computational Experiments}
\label{section:experiment}
We implement and compare our algorithms with
LZ\_OG since it has been shown to be the most
time efficient in the experiments
of~\cite{ohlebusch11:_lempel_ziv_factor_revis}.
We also implement a variant LZ\_iOG which incorporates the interleaving
optimization for $\LPF$ and $\PrevOcc$ arrays.
We have made the source codes publicly available at
\url{http://code.google.com/p/lzbg/}.

All computations were conducted on a Mac Xserve (Early 2009)
with 2 x 2.93GHz Quad Core Xeon processors and 24GB Memory,
only utilizing a single process/thread at once.
The programs were compiled using the GNU C++ compiler ({\tt g++}) 4.2.1
with the {\tt -fast} option for optimization.
The running times are measured in seconds, starting from after the
suffix array is built, and the average of 10 runs is reported.

We use the data of
\url{http://www.cas.mcmaster.ca/~bill/strings/},
used in previous work.
Table~\ref{table:data-bill} shows running times of the algorithms,
as well as some statistics of the dataset.
The running times of the fastest algorithm for each data is shown in bold.
The fastest running times for the variant that uses only $13N$ bytes 
is prefixed with `$\triangleright$'.

The results show that all the variants of our algorithms constantly
outperform LZ\_OG and even LZ\_iOG for all data tested,
and in some cases can be up to 2 to 3 times faster.
We can see that iBGS is fastest when the data is not extremely 
repetitive, and the average length of the factor is not so large, while
iBGT is fastest for such highly repetitive data. iBGT is also the
fastest when we restrict our attention to the algorithms that use only
$13N$ bytes of work space.

  \begin{sidewaystable}
  \caption{
    Running times (seconds) of algorithms and various statistics for the data set
    of \protect\url{http://www.cas.mcmaster.ca/~bill/strings/}.}
  \label{table:data-bill}
    \footnotesize
    \begin{center}
    
%%% Local Variables: 
%%% mode: japanese-latex
%%% TeX-master: t
%%% End: 

\newcommand{\wid}{0.8cm}
\newcommand{\widb}{1.3cm}
\begin{tabular}{|l|r|r|r|r|r|r|r|r|r|r|r|r|r|}
\hline
& LZ\_OG & LZ\_iOG & BGS & iBGS & BGL & iBGL & BGT & iBGT &&&&& \\ \cline{2-9}
      File Name
      & \parbox[c]{\wid}{\begin{center}$13N$\end{center}}
      & \parbox[c]{\wid}{\begin{center}$13N$\end{center}}
      & \parbox[c]{\wid}{\begin{center}$17N+4S_{max}$\end{center}}
      & \parbox[c]{\wid}{\begin{center}$17N+4S_{max}$\end{center}}
      & \parbox[c]{\wid}{\begin{center}$17N$\end{center}}
      & \parbox[c]{\wid}{\begin{center}$17N$\end{center}}
      & \parbox[c]{\wid}{\begin{center}$13N$\end{center}}
      & \parbox[c]{\wid}{\begin{center}$13N$\end{center}}
      & \parbox[c]{0.5cm}{\begin{center}$|\Sigma|$\end{center}}
      & \parbox[c]{\widb}{\begin{center}Text Size $N$\end{center}}
      & \parbox[c]{\widb}{\begin{center}\# of LZ factors \end{center}}
      & \parbox[c]{\widb}{\begin{center}Average Length of Factor\end{center}}
      & \parbox[c]{0.7cm}{\begin{center}$S_{max}$\end{center}} \\ \hline
E.coli & 0.64 & 0.58 & 0.26 & \textbf{0.23} & 0.33 & 0.29 & 0.45 & $\triangleright$ 0.37 & 4 & 4638690 & 432791 & 10.72 & 36\\ \hline
bible & 0.37 & 0.34 & 0.20 & \textbf{0.19} & 0.25 & 0.22 & 0.27 & $\triangleright$ 0.24 & 63 & 4047392 & 337558 & 11.99 & 42\\ \hline
chr19.dna4 & 10.05 & 9.25 & 4.40 & \textbf{4.00} & 5.33 & 4.71 & 7.64 & $\triangleright$ 6.54 & 4 & 63811651 & 4411679 & 14.46 & 58\\ \hline
chr22.dna4 & 5.37 & 4.91 & 2.27 & \textbf{2.06} & 2.77 & 2.44 & 4.09 & $\triangleright$ 3.45 & 4 & 34553758 & 2554184 & 13.53 & 43\\ \hline
fib\_s2178309 & 0.06 & 0.06 & 0.05 & 0.06 & 0.06 & 0.05 & 0.05 & $\triangleright$ \textbf{0.05} & 2 & 2178309 & 31 & 70268.00 & 16\\ \hline
fib\_s3524578 & 0.11 & 0.11 & 0.10 & 0.10 & 0.10 & 0.10 & 0.10 & $\triangleright$ \textbf{0.09} & 2 & 3524578 & 32 & 110143.00 & 16\\ \hline
fib\_s5702887 & 0.18 & 0.18 & 0.15 & 0.16 & 0.16 & 0.15 & 0.15 & $\triangleright$ \textbf{0.14} & 2 & 5702887 & 33 & 172815.00 & 17\\ \hline
fib\_s9227465 & 0.30 & 0.30 & 0.26 & 0.27 & 0.27 & 0.26 & 0.26 & $\triangleright$ \textbf{0.24} & 2 & 9227465 & 34 & 271396.00 & 17\\ \hline
fib\_s14930352 & 0.50 & 0.49 & 0.43 & 0.44 & 0.44 & 0.43 & 0.42 & $\triangleright$ \textbf{0.39} & 2 & 14930352 & 35 & 426581.00 & 18\\ \hline
fss9 & 0.09 & 0.08 & 0.08 & 0.08 & 0.08 & 0.08 & 0.07 & $\triangleright$ \textbf{0.07} & 2 & 2851443 & 40 & 71286.10 & 22\\ \hline
fss10 & 0.40 & 0.39 & 0.36 & 0.37 & 0.36 & 0.35 & 0.34 & $\triangleright$ \textbf{0.32} & 2 & 12078908 & 44 & 274521.00 & 24\\ \hline
howto & 4.20 & 3.91 & 2.30 & \textbf{2.15} & 2.79 & 2.51 & 3.28 & $\triangleright$ 2.91 & 197 & 39422105 & 3063929 & 12.87 & 616\\ \hline
mozilla & 5.30 & 4.95 & 3.19 & \textbf{3.13} & 3.91 & 3.65 & 4.31 & $\triangleright$ 3.86 & 256 & 51220480 & 6898100 & 7.43 & 3964\\ \hline
p1Mb & 0.08 & 0.07 & \textbf{0.05} & 0.05 & 0.06 & 0.06 & 0.05 & $\triangleright$ 0.05 & 23 & 1048576 & 216146 & 4.85 & 38\\ \hline
p2Mb & 0.23 & 0.21 & \textbf{0.11} & 0.12 & 0.15 & 0.15 & 0.17 & $\triangleright$ 0.14 & 23 & 2097152 & 406188 & 5.16 & 40\\ \hline
p4Mb & 0.58 & 0.52 & 0.26 & \textbf{0.26} & 0.35 & 0.33 & 0.43 & $\triangleright$ 0.35 & 23 & 4194304 & 791583 & 5.30 & 42\\ \hline
p8Mb & 1.27 & 1.15 & 0.55 & \textbf{0.55} & 0.73 & 0.70 & 0.94 & $\triangleright$ 0.78 & 23 & 8388608 & 1487419 & 5.64 & 898\\ \hline
p16Mb & 2.70 & 2.43 & 1.18 & \textbf{1.16} & 1.52 & 1.46 & 2.08 & $\triangleright$ 1.74 & 23 & 16777216 & 2751022 & 6.10 & 898\\ \hline
p32Mb & 5.58 & 5.02 & 2.47 & \textbf{2.44} & 3.14 & 3.03 & 4.43 & $\triangleright$ 3.74 & 24 & 33554432 & 5040051 & 6.66 & 898\\ \hline
rndA2\_4Mb & 0.49 & 0.45 & 0.20 & \textbf{0.18} & 0.24 & 0.20 & 0.35 & $\triangleright$ 0.28 & 2 & 4194304 & 201910 & 20.77 & 36\\ \hline
rndA2\_8Mb & 1.08 & 0.99 & 0.42 & \textbf{0.38} & 0.50 & 0.43 & 0.77 & $\triangleright$ 0.63 & 2 & 8388608 & 385232 & 21.78 & 37\\ \hline
rndA21\_4Mb & 0.64 & 0.58 & 0.28 & \textbf{0.28} & 0.38 & 0.37 & 0.47 & $\triangleright$ 0.37 & 21 & 4194304 & 970256 & 4.32 & 34\\ \hline
rndA21\_8Mb & 1.43 & 1.28 & 0.61 & \textbf{0.60} & 0.83 & 0.79 & 1.05 & $\triangleright$ 0.85 & 21 & 8388608 & 1835235 & 4.57 & 37\\ \hline
rndA255\_4Mb & 0.65 & 0.58 & \textbf{0.38} & 0.39 & 0.51 & 0.47 & 0.49 & $\triangleright$ 0.40 & 255 & 4194304 & 2005584 & 2.09 & 35\\ \hline
rndA255\_8Mb & 1.43 & 1.27 & 0.84 & \textbf{0.84} & 1.12 & 1.04 & 1.10 & $\triangleright$ 0.92 & 255 & 8388608 & 3817588 & 2.20 & 38\\ \hline

\end{tabular}

    \end{center}
  \end{sidewaystable}

\noindent {\bf Acknowledgments:}
We thank Dr. Simon Gog for sharing his implementation of LZ\_OG, and
Dr. Simon Puglisi for sending us the manuscripts~\cite{kempa_lz,karkkainen_lz}.

\bibliographystyle{IEEEtran}
\bibliography{ref}

\end{document}